\documentclass[10pt,twocolumn,cleanfoot,cleanhead]{asme2e}

\usepackage{empheq, amssymb}
\usepackage{lmodern}
\usepackage{amsmath}
\usepackage{amsfonts}
\usepackage{amssymb}
\usepackage{bm} 
\usepackage{txfonts}
\usepackage[T1]{fontenc}
\usepackage[utf8]{inputenc} 
\usepackage[dvipsnames]{xcolor} 
\usepackage[american]{babel}
\usepackage{graphicx}
\usepackage{epstopdf}
\usepackage[noadjust]{cite} 
\usepackage{ltxcmds}
\usepackage{mathtools}
\usepackage{subcaption}
\usepackage{multirow}
\usepackage{blindtext}
\usepackage{hyperref}
\usepackage{enumitem}
\usepackage{ifthen}
\usepackage{etoolbox}
\usepackage{cancel}
\usepackage{empheq}
\usepackage{fontawesome}
\usepackage{soul}
\usepackage{textcomp}
\usepackage{lipsum}

\definecolor{light-gray}{gray}{0.4}
\definecolor{box-gray}{gray}{1}

\hypersetup{
    unicode=false,          
    pdftoolbar=true,        
    pdfmenubar=true,        
    pdffitwindow=false,     
    pdfstartview={FitV},    
    pdftitle={Concurrent Probabilistic Control Co-Design and Layout Optimization of Wave Energy Converter Farms using Surrogate Modeling},    
    pdfauthor={Saeed Azad and Daniel R. Herber},     
    pdfsubject={},   
    pdfkeywords = {}, 
    pdfnewwindow=true,      
    colorlinks=true,
    allcolors=blue
}
\usepackage{nomencl}
\renewcommand\nomgroup[1]{%
  \item[\bfseries
  \ifstrequal{#1}{V}{ Select Variables}{%
  \ifstrequal{#1}{B}{ Subscripts}{%
  \ifstrequal{#1}{P}{ Notation}{%
  \ifstrequal{#1}{A}{ Acronyms}{}}}}]
}

\makenomenclature



\usepackage{dblfloatfix}
\usepackage{float}

\definecolor{block-gray}{gray}{0.95}

\usepackage[framemethod=TikZ]{mdframed}

\usepackage{xpatch}

\makeatletter
\xpatchcmd{\endmdframed}
  {\aftergroup\endmdf@trivlist\color@endgroup}
  {\endmdf@trivlist\color@endgroup\@doendpe}
  {}{}
\makeatother

\usepackage{fontawesome}
\usepackage{titlesec}

\newcommand{\doi}[1]{{doi:~\href{https://doi.org/#1}{\nolinkurl{#1}}}\rmFullStop}

\newcommand*{\rmFullStop}{\rmifnextchar{.}{}{}}

\makeatletter
\newcommand{\rmifnextchar}[3]{%
  \begingroup
  \ltx@LocToksA{\endgroup#2}%
  \ltx@LocToksB{\endgroup#3}%
  \ltx@ifnextchar{#1}{%
    \def\next{\the\ltx@LocToksA}%
    \afterassignment\next
    \let\scratch= %
  }{%
    \the\ltx@LocToksB
  }%
}
\makeatother

\usepackage{colortbl}

\definecolor{light-gray}{gray}{0.6}

\newcommand{\xsection}[1]{\section[#1]{\MakeUppercase{#1}}}
\usepackage{accents}

\definecolor{needcolor}{HTML}{C62828}

\usepackage{adjustbox}
\usepackage{array,booktabs}
\allowdisplaybreaks
\usepackage{units}
\newcommand{\parm}{\mathord{\color{black!33}\bullet}}%
\newcommand{\RNum}[1]{\uppercase\expandafter{\romannumeral #1\relax}}
\newcolumntype{s}{>{\columncolor[HTML]{F5F5F5}} c}
\newcommand{\MyBold}[1]{\mathbf{#1}}
\newcommand{\Rwec}{R_{\text{wec}}}            
\newcommand{\Dwec}{D_{\text{wec}}}            
\newcommand{\tRwec}{\tilde{R}_{\text{wec}}}   
\newcommand{\tDwec}{\tilde{D}_{\text{wec}}}   
\newcommand{\Rwecmax}{\bar{R}_{\text{wec}}}   
\newcommand{\Dwecmax}{\bar{D}_{\text{wec}}}   
\newcommand{\Rwecmin}{\underaccent{\bar}{R}_{\text{wec}}}   
\newcommand{\Dwecmin}{\underaccent{\bar}{D}_{\text{wec}}}   
\newcommand{\Angpq}{\theta_{pq}}              
\newcommand{\Dispq}{l_{pq}}                   
\newcommand{\tAngpq}{\tilde{\theta}_{pq}}     
\newcommand{\tDispq}{\tilde{l}_{pq}}          
\newcommand{\Angpqmax}{\bar{\theta}_{pq}}     
\newcommand{\Dispqmax}{\bar{l}_{pq}}         
\newcommand{\Nwec}{n_{\text{wec}}}     
\newcommand{\NMBE}{m_{\text{}}}      
\newcommand{\Mass}{\MyBold{M}}       
\newcommand{\Force}{\MyBold{F}}      
\newcommand{\AddedMass}{\MyBold{A}}     
\newcommand{\DampingCoeff}{\MyBold{B}}  
\newcommand{\addedmassv}{{a}}       
\newcommand{\dampingcoeffv}{{b}}    
\newcommand{\OMF}{(\omega)}             
\newcommand{\AL}{\MyBold{w}}           
\newcommand{\KPTO}{\MyBold{K}_{\text{pto}}}           
\newcommand{\BPTO}{\MyBold{B}_{\text{pto}}}           
\newcommand{\KPTOmin}{\underaccent{\bar}{\MyBold{k}}_{\text{pto}}}           
\newcommand{\BPTOmin}{\underaccent{\bar}{\MyBold{B}}_{\text{pto}}}           
\newcommand{\KPTOmax}{\bar{\MyBold{k}}_{\text{pto}}}           
\newcommand{\BPTOmax}{\bar{\MyBold{B}}_{\text{pto}}}           

\title{Concurrent Probabilistic Control Co-Design and Layout Optimization of Wave Energy Converter Farms using Surrogate Modeling}

\author{Saeed~Azad\thanks{Corresponding author, \texttt{\href{mailto:saeed.azad@colostate.edu}{saeed.azad@colostate.edu}}} 
\affiliation{
Postdoctoral Fellow\\
Department of Systems Engineering \\
Colorado State University\\
Fort Collins, CO 80523 \\
Email:~\texttt{\href{mailto:saeed.azad@colostate.edu}{saeed.azad@colostate.edu}}\\ 
}
}

\author{Daniel~R.~Herber
\affiliation{
Assistant Professor\\
Department of Systems Engineering \\
Colorado State University \\
Fort Collins, CO 80523 \\
Email:~\texttt{\href{mailto:daniel.herber@colostate.edu}{daniel.herber@colostate.edu}}
}
}

\begin{document}
\setlength{\parskip}{0pt}
\setlength{\parsep}{0pt}
\setlength{\headsep}{0pt}
\setlength{\topsep}{0pt}

\abovedisplayshortskip=3pt
\belowdisplayshortskip=3pt
\abovedisplayskip=3pt
\belowdisplayskip=3pt

\titlespacing*{\section}{0pt}{18pt plus 1pt minus 1pt}{3pt plus 0.5pt minus 0.5pt}

\titlespacing*{\subsection}{0pt}{9pt plus 1pt minus 0.5pt}{1pt plus 0.5pt minus 0.5pt}

\titlespacing*{\subsubsection}{0pt}{9pt plus 1pt minus 0.5pt}{1pt plus 0.5pt minus 0.5pt}

\maketitle

\begin{abstract}\noindent%
\textit{Wave energy converters (WECs) are a promising candidate for meeting the increasing energy demands of today's society.
It is known that the sizing and power take-off (PTO) control of WEC devices have a major impact on their performance.
In addition, to improve power generation, WECs must be optimally deployed within a farm.
While such individual aspects have been investigated for various WECs, potential improvements may be attained by leveraging an integrated, system-level design approach that considers all of these aspects. 
However, the computational complexity of estimating the hydrodynamic interaction effects significantly increases for large numbers of WECs. 
In this article, we undertake this challenge by developing data-driven surrogate models using artificial neural networks and the principles of many-body expansion.
The effectiveness of this approach is demonstrated by solving a concurrent plant (i.e., sizing), control (i.e., PTO parameters), and layout optimization of heaving cylinder WEC devices.
WEC dynamics were modeled in the frequency domain, subject to probabilistic incident waves with farms of $3$, $5$, $7$, and $10$ WECs.
The results indicate promising directions toward a practical framework for array design investigations with more tractable computational demands.%
}%
\end{abstract}%

\vspace{1ex}
\noindent Keywords:~surrogate modeling; control co-design; layout optimization; wave energy converter farms; energy systems

\xsection{Introduction}\label{sec:introduction}

Wave energy, with its temporal and spatial availability, low variability, and high predictability, is a promising source of renewable energy \cite{ning2022modelling}.
However, its technology readiness level (TRL), which is often used to classify the development maturity of a new technology, is still low compared to wind and solar technologies \cite{straub2015search}, indicating that more investment and research are required to improve their techno-economic performance \cite{tan2021influence}. 

Among the many approaches undertaken as a part of this effort, the primary focus has been on the sizing of the device (i.e.,~plant) and/or its power take-off (PTO) (i.e.,~control) \cite{mccabe2022multidisciplinary, neshat2020new, herber2013wave, tan2022application, coe2020initial}.
Since the economic viability of wave energy converter (WEC) devices depends strongly on the energy generated, WECs must be carefully deployed in an array.
While WEC arrays reduce the installation, maintenance, and operation costs \cite{abdulkadir2023optimization}, they result in a complex hydrodynamic interaction effect that appears mainly due to the presence of multiple WECs in close proximity \cite{falnes1980radiation}.   
This interaction effect can be constructive or destructive \cite{borgarino2012impact}.
Therefore, the spatial configuration of WECs within an array must be carefully selected to ensure a constructive effect.
This exploration is generally done through array or layout optimization of WECs \cite{abdulkadir2023optimization, neshat2022layout, mercade2017layout}.

As emphasized in Ref.~\cite{ringwood2023empowering}, potential improvements in WEC array performance may be realized by leveraging a system-level design framework that considers plant, control, and layout concurrently.   
However, accurate estimation of the hydrodynamic interaction effect in array optimization in general, and this integrated design framework in particular, is computationally expensive \cite{lyu2019optimization}.
For example, a single call to the boundary element method (BEM) solver \texttt{Nemoh} for a two- and three-WEC farm using axisymmetric meshes takes about $384~\textrm{s}$, and $1305~\textrm{s}$, respectively. 
Because of this cost, the design of WEC farms using accurate numerical solutions has been generally limited in the complexity of the array (fixed array geometries such as square and triangular \cite{borgarino2012impact}), and the total number of WECs in the farm \cite{lyu2019optimization}.

In this article, we estimate the complex hydrodynamic interaction effect, up to second-order, by constructing data-driven surrogate models using artificial neural networks (ANNs), along with a hierarchical interaction decomposition approach inspired by concepts from many-body expansion (MBE) \cite{gordon2012fragmentation, zhang2020surrogate}.
Since BEM solvers are known to provide the accurate hydrodynamics \cite{babarit2013park}, the ANN models were trained on data generated from an open-source, BEM solver \texttt{Nemoh} \cite{babarit2015theoretical, Kurnia2022, kurnia2022second}, for one- and two-WEC studies, leading to a second-order approximation of the complex hydrodynamic interaction effect. 
These surrogate models are then utilized within the optimization framework to implement a surrogate-assisted optimization approach that has the potential to efficiently address the issue of accurate estimation of hydrodynamic coefficients within large arrays.
Through this, we aim to develop a practical framework that enables the design of more complex WEC arrays compared to other similar works in the literature with fewer assumptions \cite{lyu2019optimization}.

This article presents some preliminary results towards the concurrent optimization of plant (uniform across the farm), control, and layout for heaving cylinder WEC devices within a farm. 
This framework naturally combines an stochastic in expectation uncertain control co-design (SE-UCCD) \cite{azad2022control} formulation with the layout optimization problem.
By enabling the simultaneous implementation of an integrated design approach, known as control co-design (CCD) \cite{garcia2019control, strofer2023control, coe2020initial, o2017co,pena2022control} with layout optimization for half-submerged cylindrical, heaving WEC devices in the presence of uncertainties from incident waves, this article attempts to pave the way for more complex investigations of WEC farms.  

The heaving WEC is characterized as a point absorber, i.e., having a relatively small dimension with respect to the prevailing wavelength \cite{ning2022modelling}.
This device is modeled in the frequency domain, with a PTO system that exerts a load force on the oscillating body while storing energy (thereby providing a bi-directional power flow through a reactive control strategy) \cite{ning2022modelling}. 
To calculate the absorbed power, the probability distribution of waves is constructed based on historical data collected from Pacific Islands in close proximity to the Hawaiian Islands over the lifetime of the device.
This data set is openly accessible from Refs.~\cite{storlazzi2015future, erikson2016wave}.
The optimization problem is formulated with the objective function of maximizing the expected value of power over the lifetime of the farm per unit volume of the device.

The remainder of this article is organized as follows:
Sec.~\ref{sec:section2} presents the methodological discussion, including probabilistic wave modeling, wave-structure interactions, dynamics and equations of motions for WECs, array geometry and considerations, surrogate modeling with an emphasis on ANN, and a brief introduction into principles of MBE;
Sec.~\ref{sec:Constructing_Surrogate_Models} describes the construction and validation of surrogate models using ANNs and the BEM solver \texttt{Nemoh};
Sec.~\ref{sec:UCCDandLayoutOpt} starts with introducing the concurrent UCCD and layout optimization problem and then presents results associated with multiple array investigations;
Finally, Sec.~\ref{sec:conclusion} presents conclusions and limitations of the current study, as well as potential future work.

\xsection{Methods}\label{sec:section2}

In this section, we start by describing the wave climate and its modeling considerations. 
Calculation of hydrodynamic coefficients, along with the basics of potential flow theory and boundary element method in \texttt{Nemoh} are presented next. 
Then equations of motion for a heaving cylinder WEC are introduced in the frequency domain, and some considerations regarding layout geometry are presented. 
The section concludes with a discussion on surrogate models and many-body expansion principles.    
\begin{figure}
    \centering
    \includegraphics[width=\columnwidth]{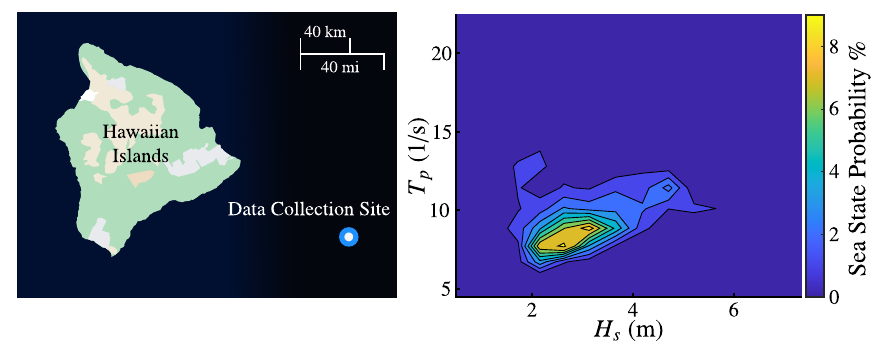}
    \caption{Data collection site and the wave scatter diagram created using the joint probability distribution of significant wave heights and wave periods.}
    \label{fig:Wave_Climate}
\end{figure}

\subsection{Wave Climate and Modeling} \label{subsec:WaveClimate}

In wave energy applications, wind-generated gravity surface waves are of interest.
These are waves resulting from the wind blowing over the ocean surface, dominated by gravity and inertial forces \cite{ning2022modelling}.
The site of interest is located in the Pacific Islands, off the coast of the Hawaiian Islands.
$30$ years of historical data, collected from $1976$ through $2005$, is used to estimate the joint probability distribution of waves for various significant wave heights $H_{s}$ and wave periods $T_{p}$. 
Since the resulting data form a non-parametric representation of the probability density function, a kernel distribution characterized by a smoothing function and a bandwidth value was utilized.
Using a Gaussian quadrature approach with $n_{gq}$ points in each dimension, we first obtained the Legendre-Gauss nodes and weights.
The probability distribution of waves was then constructed using these points in \texttt{Matlab}'s \textit{ksdensity} function, resulting in the annual probability matrix. 
Figure~\ref{fig:Wave_Climate} presents the location of the site, along with the wave contour diagram for year $1$ of the study.

The sea state is described using the JOint North Sea WAve Project (JONSWAP) spectrum, defined as:
\begin{align}
    \label{eqn:JS}
    S_{JS}(H_{s},T_{p},\omega) = \alpha_{s} \omega^{-5} \exp{\left[ -\beta_{s} \omega^{-4}\right]} 
\end{align}
\noindent
where $\omega$ is the angular frequency, and $\alpha_{s}$ and $\beta_{s}$ are parameters of the spectrum defined as:
\begin{align}
    \label{eqn:JS_parameters_AlphaBeta}
    \alpha_{s} &= \dfrac{\beta_{s}}{4}H_{s}^{2}C(\gamma)\gamma^{r}\\ 
    \beta_{s} &=\dfrac{5}{4}\omega_{p}^{4}   
\end{align}
\noindent
where $\omega_{p}$ is the peak angular frequency, and $C(\cdot)$ is a normalizing factor calculated as:
\begin{align}
    \label{eqn:JS_parameters_C}
    C(\gamma) &= 1 - 0.287 \ln(\gamma) 
\end{align}
\noindent
In these equations, $\gamma$ is defined as:
\begin{align}
    \label{eqn:JS_parameters_gamma}
    \gamma = 
    \begin{cases}
         5~ &\textrm{for}~~ \dfrac{T_{p}}{\sqrt{H_{s}}} \leq 3.6\\
         \exp{(5.75 -1.15\dfrac{T_{p}}{\sqrt{H_{s}}})}  &\textrm{for}~~ 3.6 \leq \dfrac{T_{p}}{\sqrt{H_{s}}} \leq 5 \\
         1 &\textrm{for}~~ \dfrac{T_{p}}{\sqrt{H_{s}}} > 5\\
    \end{cases}
\end{align}
\noindent
Finally, $r$ is defined as:
\begin{align}
    \label{eqn:JS_parameters_r}
    r = \exp{\left[\dfrac{-1}{2\sigma^{2}}\left(\dfrac{\omega}{\omega_{p}} - 1\right)^2\right]} \quad 
    \textrm{where}~\sigma = 
    \begin{cases} 
        0.07~ \textrm{for} ~\omega \leq \omega_{p}\\
        0.09~ \textrm{for}~ \omega > \omega_{p}
    \end{cases}
\end{align}
\noindent
For more details on this spectrum and its associated parameters, the readers are referred to Ref.~\cite[pp.~71]{ning2022modelling}.

\begin{figure}[t]
    \captionsetup[subfigure]{justification=centering}
    \centering
    \begin{subfigure}{1\columnwidth}
    \centering
    \includegraphics[scale=1]{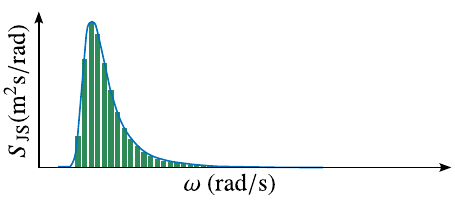}
    \caption{JONSWAP spectrum.}
    \label{subfig:spectrum}
    \end{subfigure}
    \begin{subfigure}{1\columnwidth}
    \centering
    \includegraphics[scale=1]{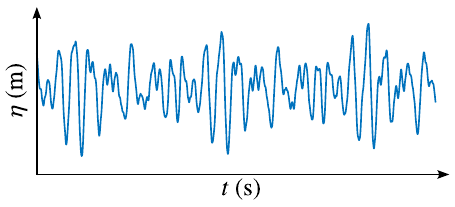}
    \caption{Irregular wave signal resulting from superposition of harmonic waves.}
    \label{subfig:IrrWave}
    \end{subfigure}%
    \captionsetup[figure]{justification=centering}
    \caption{JONSWAP spectrum and {an irregular wave signal created using $200$ harmonic waves}.}
    \label{fig:WaveSpectrum}
\end{figure}

Using this spectrum, the incident wave field can be modeled by irregular waves constructed through the superposition of $n_{r}$ regular waves.
Assuming that the angle of wave direction $\beta_{w}$ is $0$, the irregular incident wave can be approximated as: 
\begin{align}
    \label{eqn:irregularwave}
    \eta (x,t) = \sum_{i=1}^{n_{r}} \dfrac{H_{i}}{2}\cos{(k_{i}x - \omega_{i} t + \phi_{i})}
\end{align}
\noindent
where $H_{i}$ is the wave amplitude, $\theta$ is the randomly generated wave phase, and $k$ is the wave number satisfying the dispersion relation:
\begin{align}
    \label{eqn:dispersion}
    \omega^{2} = gk\tanh{kh} \approx 
    \begin{cases}
        gk~ &\textrm{as}~~ kh \to \infty\\
        gk^{2}h &\textrm{as}~~ kh \to 0
    \end{cases}
\end{align}
\noindent
where $g$ is the gravitational acceleration, and $h$ is the water depth.
Figure~(\ref{fig:WaveSpectrum}) shows the JONSWAP spectrum and a wave signal generated for an arbitrary $H_{s}$ and $T_{p}$ associated with the site of study. 

\subsection{Wave-Structure Interactions} 
\label{subsec:Wave_Structure_Interactions}

Before discussing the design and dynamics of WEC devices, it is necessary to understand the ocean waves and wave-structure interactions.
This section describes some of the basic principles without going into too much of the mathematical description of such concepts.
Interested readers may refer to Refs.~\cite{ning2022modelling, folley2016numerical, falnes2020ocean} for more details. 

\subsubsection{Linear Potential Flow Theory.}
\label{subsubsec:Linear_Potential_Flow_Theory}
~In their most fundamental form, the ocean waves may be described through Navier-Stokes equations, which along with continuity equation and incompressibility constraint, result in a system of nonlinear partial differential equations \cite{ning2022modelling}.
The numerical solution of this system of equations through computational fluid dynamics (CFD) is computationally expensive. 
However, by making simplifying assumptions, such as incompressible, inviscid (negligible viscosity), and irrotational flow, one can utilize potential flow theory, which is characterized by the Laplace equation, and boundary constraints at the free surface and any rigid boundary.
By further assuming non-steep waves (i.e.,~small wave height in relation to wave length), the free-surface boundary constraints can be linearized, resulting in linear potential theory, also known as Airy wave theory \cite{ning2022modelling}.  

\subsubsection{Hydrodynamic and Hydrostatic Forces.}
~In linear potential flow theory, the fluid velocity potential $\phi$ is divided into potentials corresponding to incident $\phi_{i}$, scattered $\phi_{s}$, and radiated waves $\phi_{r}$:
\begin{align}
    \phi = \phi_{i} + \phi_{s} + \phi_{r}
\end{align}
Incident waves represent the propagation of the wave in the absence of any structure.
Scattered waves appear as a result of the interaction of the waves and the motionless structures, while radiated waves result from the motion of the structure.

The force and force moment on a structure is separated into forces resulting from the incident and scattered waves, also known as the excitation force and the radiation force.
The excitation force is created by the action of incident waves on motionless structures.
Therefore, the excitation force captures two effects: the motion of the incident wave, also known as the Froude-Krylov force, as well as the scattering effect. 
The radiation force characterizes the forces applied to the structure as a result of its own oscillatory motion in the absence of an incident wave field.
The real part of the radiation force constitutes what is known as the added mass, while the imaginary part is the damping coefficients.

The forces created by scattered and radiated waves are of significant importance in the layout optimization of WEC devices because, as opposed to the incident wave that travels only in one direction, the scattered and radiated waves propagate along all directions, thereby affecting all WEC devices, regardless of their location.
Therefore, they redistribute part of the incident energy in all directions \cite{babarit2013park}.
Because of this phenomenon, the design, control, and layout optimization of WECs have strong interactions and must be approached concurrently.  
Hydrostatic forces result from the change in the hydrostatic pressure on the wetted surface of the body as it moves from its equilibrium position.
Analytical and numerical solutions exist to calculate these forces. 
In the following section, we briefly describe one such numerical method known as the boundary element method or BEM.

\subsubsection{Boundary Element Method in Nemoh.}
~In BEM, systems of partial differential equations from linear potential theory are formulated into the boundary integral form and transformed into a problem of source distribution on the body surface using Green's theorem \cite{penalba2017using}.   
While many BEM solvers are available, in this study we use Nemoh, an open-source frequency-domain solver.
To use \texttt{Nemoh,} the user must provide the mesh for the desired body.
Using the \texttt{Matlab} wrapper provided in the software package, a symmetric mesh was generated for the cylindrical WEC device based on the specified radius and draft dimensions and used in \texttt{Nemoh}.

\subsection{Dynamics and Control of WECs} 
\label{subsec:Dynamics_of_WEC_Arrays}
Despite high nonlinearities, the hydrodynamic interactions between a WEC device and ocean waves can be simplified when the device exhibits small amplitudes in its oscillatory motions.
This, along with a linear representation of forces involved in WEC dynamics, can be utilized to carry out the modeling in the frequency domain.
These frequency-domain models are perhaps the first design space that engineers thoroughly investigate to gain early-stage insights before moving on to the more expensive time-domain models.
Due to their computational efficiency, these models are suitable for the preliminary assessment of a system-level design framework in which farm layout design considerations are accompanied by control co-design efforts. 
In this section, we briefly present the relevant formulations for the dynamics of WEC arrays in the frequency domain.

Using linear potential flow theory, and considering regular waves with radial frequency of $\omega$ and unit amplitude as an input, the equation of motion for $\Nwec$ buoys in the frequency domain can be described as:
\begin{align}
\label{eqn:EquationofMotion1}
     -{\omega}^{2} \Mass \hat{\bm{\xi}}\OMF = \hat{\Force}_{\text{FK}}\OMF + \hat{\Force}_{\text{s}}\OMF + \hat{\Force}_{\text{r}}\OMF +  \hat{\Force}_{\text{hs}}\OMF + \hat{\Force}_{\text{pto}}\OMF
\end{align}
\noindent
where $\hat{\parm}$ is the complex amplitude of $\parm$, $\hat{\bm{\xi}}(\cdot) \in \mathbb{R}^{\Nwec \times 1}$ is the displacement vector, $\hat{\Force}_{\text{FK}}(\cdot)$ is the Froude-Krylov force, $\hat{\Force}_{\text{s}}(\cdot)$ is the scattering force vector, $\hat{\Force}_{\text{r}}(\cdot)$ is the radiation force, $ \hat{\Force}_{\text{hs}}(\cdot)$ is the hydrostatic force, $\hat{\Force}_{\text{pto}}(\cdot)$ is the power-take-off (PTO) force, all defined in $\in \mathbb{R}^{\Nwec \times 1}$.
In addition, $\Mass \in \mathbb{R}^{\Nwec \times \Nwec}$ is the diagonal mass matrix.
The excitation force is defined as the sum of Froude-Krylov and scattering forces:
\begin{align}
\label{eqn:F_excitation}
    \hat{\Force}_{\text{e}}\OMF = \hat{\Force}_{\text{FK}}\OMF + \hat{\Force}_{\text{s}}\OMF
\end{align}
\noindent
The radiation force is calculated as a function of hydrodynamic damping coefficient matrix $\DampingCoeff(\cdot)$, and added mass coefficient matrix $\AddedMass(\cdot)$ as:
\begin{align}
\label{eqn:F_radiation}
    \hat{\Force}_{\text{r}}\OMF = -i{\omega} \DampingCoeff\OMF\hat{\bm{\xi}}\OMF + {\omega}^{2}\AddedMass\OMF\hat{\xi}\OMF
\end{align}
\noindent
where $\DampingCoeff(\cdot) \in \mathbb{R}^{\Nwec \times \Nwec}$ captures the dissipated energy transmitted from WEC motions to the water (propagating away from the body dissipative effect), and $\AddedMass(\cdot)\in \mathbb{R}^{\Nwec \times \Nwec}$ represents the inertial increase due to water displacement as a result of the WEC motion (reactive effect) \cite{folley2016numerical}. 
Considering $\Nwec$ WEC devices in a fixed array with only heave motion, $\AddedMass\OMF$ and $\DampingCoeff\OMF$ matrices have the following form \cite{lyu2019optimization}:
\begin{align}
    \label{eqn:AandBMatrices}
    \AddedMass\OMF &= \begin{bmatrix}
        \addedmassv_{11}\OMF&~~ \addedmassv_{12}\OMF&~~ \addedmassv_{13}\OMF&~~ \cdots&~~ \addedmassv_{1\Nwec}\OMF&\\
        \addedmassv_{21}\OMF&~~ \addedmassv_{22}\OMF&~~ \addedmassv_{23}\OMF&~~ \cdots&~~ \addedmassv_{2\Nwec}\OMF&\\
        \vdots& \vdots&  \vdots& \vdots& \vdots \\
        \addedmassv_{\Nwec1}\OMF& \cdots& \cdots& \cdots& \addedmassv_{\Nwec\Nwec}\OMF&
    \end{bmatrix}\\
        \DampingCoeff\OMF &= \begin{bmatrix}
        \dampingcoeffv_{11}\OMF&~~ \dampingcoeffv_{12}\OMF&~~ \dampingcoeffv_{13}\OMF&~~ \cdots&~~ \dampingcoeffv_{1\Nwec}\OMF&\\
        \dampingcoeffv_{21}\OMF&~~ \dampingcoeffv_{22}\OMF&~~ \dampingcoeffv_{23}\OMF&~~ \cdots&~~ \dampingcoeffv_{2\Nwec}\OMF&\\
        \vdots& \vdots& \vdots& \vdots& \vdots \\
        \dampingcoeffv_{\Nwec1}\OMF& \cdots& \cdots& \cdots& \dampingcoeffv_{\Nwec\Nwec}\OMF&
    \end{bmatrix}
\end{align}
The hydrostatic force, which results from the balance between buoyancy and gravity is calculated as:
\begin{align}
    \hat{\Force}_{\text{hs}}\OMF = - G \hat{\bm{\xi}}\OMF  
\end{align}
\noindent
where $G$ is the hydrostatic coefficient, calculate as $G = \rho gS$, with $S$ being the cross-sectional area at the undisturbed sea level calculated as a function of the WEC radius: $ \pi \Rwec^2$ for a heaving cylinder. 

At the core of WEC device functionality is a PTO system that converts mechanical motion into electricity. 
PTO design has significant implications on the sizing and economic performance of WECs \cite{tan2021influence, tan2020feasibility}. Therefore it is necessary to investigate the PTO system from the early stages of the design process.
It is well known within the WEC literature that incorporating active control strategies significantly increases the energy generated through WECs \cite{herber2013wave, tedeschi2010analysis, clement2012discrete}.
Reactive control is one of the earliest control strategies developed for WEC devices.
It enables a bidirectional power flow between the PTO spring and the buoy \cite{ning2022modelling}.
In its linear form, the PTO force is composed of two contributions:
\begin{align}
    \hat{\Force}_{\text{pto}}({\omega}) = -i{\omega}\BPTO\hat{\bm{\xi}}{(\omega)} - \KPTO \hat{\bm{\xi}}{(\omega)}
\end{align}
\noindent
where $\KPTO \in \mathbb{R}^{\Nwec \times \Nwec}$ and $\BPTO \in \mathbb{R}^{\Nwec \times \Nwec}$ are diagonal matrices of stiffness and damping of PTO systems for WEC devices, respectively.
The complex amplitude of the motions of all WECs for a regular wave of frequency $\omega$ and unit amplitude can now be described as a transfer function matrix:
\begin{align}
{\hat{\bm{\xi}}(\omega)} &= {\mathbf{H}(\omega)\hat{\Force}_{\text{e}}({\omega})} \label{eq:TF} \\
\mathbf{H}(\omega) &= \left [ [{\omega}^2(\Mass+\AddedMass({\omega})) + G + \KPTO] + i{\omega}(\DampingCoeff({\omega}) + \BPTO)  \right ]^{-1} \label{eq:TF_denom}
\end{align}
From here, captor velocity and acceleration can be calculated as:
\begin{align}
    \hat{\dot{\bm{\xi}}} &= i{\omega}\hat{\bm{\xi}}\\
    \hat{\ddot{\bm{\xi}}} & = - {\omega}^2\hat{\bm{\xi}}
\end{align}
\noindent
These complex amplitudes can be used to represent the displacement, velocity, and acceleration in time through the following equations:
\begin{align}
    \bm{\xi}(t) &= \text{Re}\{ \hat{\bm{\xi}}({\omega})\exp(i{\omega}t)\}\\
    \dot{\bm{\xi}}(t) &= \text{Re}\{ i{\omega}\hat{\bm{\xi}}({\omega})\exp(i{\omega}t)\}\\
    \ddot{\bm{\xi}}(t) &= \text{Re}\{ - {\omega}^2 \hat{\bm{\xi}}({\omega})\exp(i{\omega}t)\}
\end{align}
\noindent
The time-averaged absorbed mechanical power for a sea state with significant wave height of $H_{s}$ and peak period of $T_{p}$ can then be described as:
\begin{align}
    {\mathbf{p}_{m}(H_{s}, T_{p}, \omega) = \frac{1}{2}{\omega}^2 \hat{\bm{\xi}}^{T}\BPTO \hat{\bm{\xi}} }  \label{eq:pto_power1}
\end{align}
\noindent
To calculate the absorbed power for the site of interest for year $y$, it is necessary to estimate the power production of the wave farm in each desired sea state by integrating the product of the wave spectrum with the time-averaged power of Eq.~(\ref{eq:pto_power1}) over all frequencies \cite{neshat2022layout, borgarino2012impact}:
\begin{align}
    {\mathbf{p}_{i}(H_{s}, T_{p}, y) = \int_{0}^{\infty}2S_{JS}(H_{s},T_{p},\omega)\mathbf{p}_{m}(H_{s}, T_{p}, \omega)d\omega } \label{eq:pto_power2}
\end{align}
\noindent
where $\mathbf{p}_{i}(H_{s}, T_{p}, y)$ is the mechanical power matrix.
Considering all sea states (which are now discretized by $n_{gq}$ Gauss quadrature points), this equation can be estimated using \cite{mercade2017layout}:
\begin{align}
    {\mathbf{p}_{i}(H_{s}, T_{p}, y)} &= {\sum_{k = 0}^{n_{w}} 2 \Delta\omega_{k} S_{JS}(H_{s},T_{p},\omega_{k})\mathbf{p}_{m}(H_{s}, T_{p}, \omega_{k})}  \label{eq:pto_power3} 
\end{align}
\noindent
where $n_{w}$ is the number of frequencies in the discretized form.
Now, considering the number of years in the study, where $y=1,2, \dots n_{yr}$ and their associated probability matrices, the average power can be calculated as \cite{neary2014methodology}:
\begin{align}
    {p_{a} = \eta_{\text{pcc}}\eta_{\text{oa}}\eta_{\text{t}}\sum_{y = 1}^{n_{yr}}\mathbf{p}_{i}(H_{s}, T_{p}, y) \mathbf{p}_{r}(H_{s},T_{p}, y)}
\end{align}
\noindent
where $p_{a}$ is the average power.
The joint probability distribution of the wave climate in the $y$th year of the study is described by $\mathbf{p}_{r}(H_{s},T_{p}, y)$.
In this equation, $\eta_{\text{pcc}}$ is the efficiency of the power conversion chain, $\eta_{\text{oa}}$ is the operational availability, and $\eta_{\text{t}}$ is the transmission efficiency.
Ideally, maximizing the ratio between the average power and the total cost is desired, including development, infrastructure, PTO, maintenance, etc. 
In this article, however, we use a much simpler objective function that considers the average power per unit volume of the wave-absorbing body, which is expressed as \cite{falnes2020ocean}:
\begin{align}
    p_{v} = \frac{p_{a}}{\pi \Rwec^2 \Dwec} 
\end{align}
\noindent
where $p_{v}$ is the average power per unit volume of the device.
$\Rwec$ and $\Dwec$ are the radius and draft of the heaving cylinder WEC device, respectively.

\subsection{Array Considerations} 
\label{subsec:ArrayConsiderations}

\begin{figure}
    \centering
    \includegraphics[scale=0.9]{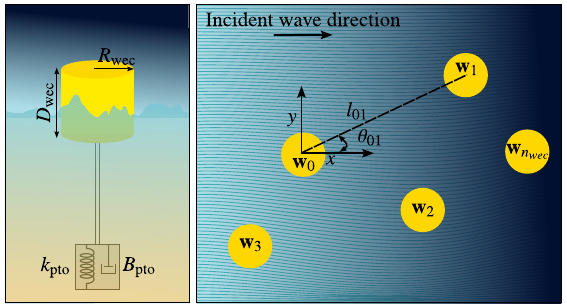}
    \caption{Illustration of WEC and its layout considerations.}
    \label{fig:WaveFarm}
\end{figure}

In this study, we consider an array with a total of $\Nwec$ WEC devices that are fully characterized by the $2-$by-$\Nwec$ dimensional layout matrix $\AL = [\mathbf{w}_{1}, \mathbf{w}_{2}, \cdots, \mathbf{w}_{\Nwec}]$.
Each element of $\Nwec$ is a vector composed of the center coordinates of each body in the Cartesian coordinate system, such that the $p$th body is characterized by  $\mathbf{w}_{p} = [x_{p}, y_{p}]^{T}$.
These bodies are subject to an incident wave propagating along the direction with angle $\beta_{w}$ relative to the $x$-axis.
Since the coordinate system may be rotated for these axisymmetric bodies, the incident wave angle may be set to $0$ without any loss of generality.
We further characterize the relative distance between two WECs, namely $p$th and $q$th bodies, and their relative angle with respect to the positive direction of the $x$-axis as $\Dispq$ and $\Angpq$, respectively.
Figure \ref{fig:WaveFarm} illustrates WECs in an array.

\subsection{Surrogate Models for Hydrodynamic Interactions}
\label{subsec:SM_HIE}

Surrogate models provide a valuable tool for a simplified approximation of input/output relationships for computationally expensive tasks.  
The usage of surrogate models in general, and ANNs in particular in the WEC literature, is not new.
Anderlini et al. used ANNs to map the significant wave height, wave energy period, PTO damping, and its stiffness to the mean absorbed power in Ref.~\cite{anderlini2017reactive}.
A shallow neural network machine learning algorithm was used in Ref.~\cite{thomas2018experimental} to find the optimal latching times for the latching control of WECs.
An ANN was developed in Ref.~\cite{valerio2008identification} for identification and control of a fully-submerged WEC.
Reference~\cite{zhang2020surrogate} developed surrogate models to capture the hydrodynamic interactions effect within a WEC farm using Gaussian process regression. 
In this work, we follow the path of Ref.~\cite{zhang2020surrogate} to develop surrogate models that can estimate the hydrodynamic interaction effect using hierarchical interaction decomposition and MBE principles with ANNs.

\subsubsection{Artificial Neural Networks.}
\label{subsec:ANN}
~ANN is a model comprised of an interconnected network of nodes or neurons in a layered structure. 
Each neuron uses a transfer function (such as the logistic function) to output a value between $0$ and $1$ depending on the weighted sum of its inputs. 
These neurons are characterized by a weight term that defines the input/output function of the network \cite{lecun2015deep}.  
These weights, along with the network structure, may be adjusted to ensure that ANN performs reasonably well in predicting the outputs. 
These weights are found by fitting or learning on the training data through an algorithm that computes the gradient vector to minimize a prediction error.
A performance measure, such as the mean-squared error, characterizes the error.  
For more information about ANN, the readers are referred to Ref.~\cite{haykin1998neural}.

In this article, we utilize feedforward shallow neural networks.
In such networks, the neurons in each hidden layer use the outputs of all nodes from the previous layer as input.
This structure allows the network to capture complex nonlinear relationships in the data. 
Calibrating the ANN may take some effort; however, the resulting surrogate models can predict the desirable outputs with high accuracy and, thus, facilitate the effective implementation of layout optimization for WECs. 

\subsubsection{Many-body Expansion.}
\label{subsec:MBE}
~Estimating the interaction effect among a large number of bodies is a problem relevant to a multitude of disciplines, including quantum mechanics \cite{lee1957many} and molecular dynamics \cite{gordon2012fragmentation}. 
Many-body expansion method (MBE) has been developed to estimate the total interaction effect among $\Nwec$ bodies as the summation of effects corresponding to a finite number of clusters.
These clusters are systematically selected to capture the effects of a single, two-, three-, and $\NMBE$-bodies \cite{suarez2009thermochemical}. 
Representing the desired hydrodynamic interaction effect with $\psi(\AL)$, MBE up to $\NMBE$ clusters can be estimated as \cite{zhang2020surrogate}:
\begin{align}
    \label{eqn:MBE1}
    \psi(\AL) &\approx \sum_{i=1}^{\Nwec} \psi(\MyBold{w}_{i}) + \sum_{i=1}^{\Nwec-1} \sum_{j>i}^{\Nwec}\Delta\psi(\MyBold{w}_{i},\bm{w}_{j})  + \dots \\
    & \quad + \sum_{i=1}^{\Nwec-2}\sum_{j>i}^{\Nwec-1}\sum_{k>j}^{\Nwec} \Delta\psi(\MyBold{w}_{i},\MyBold{w}_{j}, \MyBold{w}_{k}) + \dots \notag \\
    & \quad + \sum_{i=1}^{\Nwec-\NMBE}\dots\sum_{k>j}^{\Nwec} \Delta\psi(\MyBold{w}_{i}, \dots, \MyBold{w}_{k}) \notag
\end{align}
\noindent
Using an alternative notation $\{ \AL\}_{l}^{\NMBE}$ to represent the $l$th distinct $m$-body cluster, this formulation can be written as:
\begin{align}
\label{eqn:MBE2}
    \psi(\AL) &\approx \sum_{l=1}^{\Nwec} \psi(\{ \AL_{l}^{1}\}) + \sum_{l=1}^{\Nwec!/[2!(\Nwec-2)!]} \Delta\psi(\{ \AL\}_{l}^{2}\})  + \dots \\
    & \quad + \sum_{l=1}^{\Nwec!/[3!(\Nwec-3)!]} \Delta\psi(\{ \AL\}_{l}^{3}\}) + \dots \notag\\
    & \quad + \sum_{l=1}^{\Nwec!/[\NMBE!(\Nwec-\NMBE)!]} \Delta\psi(\{ \AL\}_{l}^{\NMBE}\}) \notag
\end{align}
\noindent
where $\Delta\psi(\cdot)$ is the additive interaction effect, calculated as:
\begin{align}
\label{eqn:MBEadditive}
    \Delta\psi(\{\AL\}_{l}^{\NMBE}) &= \psi(\{\AL\}_{l}^{\NMBE}) - \left [ \sum_{r=1}^{\NMBE} \psi(\{ \{ \AL\}_{l}^{\NMBE} \}_{r}^{1})  \right.  + \dots \\
    &\left. + \sum_{r=1}^{\NMBE!/[2!(\NMBE-2)!]} \Delta\psi(\{ \{ \AL\}_{l}^{\NMBE} \}_{r}^{2}) + \dots + \sum_{r=1}^{\NMBE} \Delta\psi(\{ \{ \AL\}_{l}^{\NMBE} \}_{r}^{\NMBE-1}) \right] \notag
\end{align}
\noindent
In this study, we only consider interaction effects of two-body clusters. 
Therefore, for each pair of $p$ and $q$ WECs, the additive effect can be written as:
\begin{align}
\label{eqn:MBEadditive2}
    \Delta\psi(\MyBold{w}_{p}, \MyBold{w}_{q}) &= \psi(\MyBold{w}_{p}, \MyBold{w}_{q}) - \psi(\MyBold{w}_{p})-\psi(\MyBold{w}_{q})
\end{align}
\noindent

\xsection{Constructing Surrogate Models}
\label{sec:Constructing_Surrogate_Models}

In this section, we discuss some considerations in developing ANNs for hydrodynamic interaction effects and demonstrate their capability to estimate the hydrodynamic coefficients with reasonable accuracy.

\subsection{Developing Surrogate Models}
\label{subsec:Developing_Surrogate_Models}

In order to generate the training data for ANNs, the first step is to identify inputs and outputs.
In this article, we are interested in estimating radiation and excitation forces exerted on the WEC for all one- and two-WEC clusters. 
This results in an output structure with components associated with the added mass, damping coefficient, and the real and imaginary parts of the excitation force.
These quantities of interest (QoI) constitute the output vector of our ANNs and are normalized according to the following relationships \cite{zhang2020surrogate, mavrakos1987hydrodynamic, herber2013wave}:
\begin{align}
\tilde{\Force}_{\text{e}} &= \hat{\Force}_{\text{e}} /(\rho g \pi\Rwec^2 \Dwec) \\
\tilde{\AddedMass} &= \AddedMass/(\rho \pi \Rwec^2 \Dwec)\\
\tilde{\DampingCoeff} &= \DampingCoeff / (\omega \rho \pi \Rwec^2 \Dwec).
\end{align}
\noindent
Although it is theoretically possible to develop a single ANN with multiple inputs and outputs, we use a single ANN for each of these outputs in this work.
This separation helps with simplifying the task of tuning network parameters.

For the calculation of quantities of interest in the single-body cluster, a sufficient number of samples for different WEC radii and draft dimensions must be provided only at the center of the coordinate system, i.e.,~$\MyBold{w}_{0} =[0,0]^{T}$.
This is because the radiation output from these surrogate models is independent of the location of the WEC (i.e.,~$\AddedMass(\AL) = \AddedMass([0, 0]^{T})$ and $\DampingCoeff(\AL) = \DampingCoeff([0, 0]^{T})$), however, it depends on plant specifications, such as WEC radius and draft.
The excitation force output is invariant with respect to translations along the $y$-axis and is symmetric with respect to the wave propagation direction (i.e., $x$-axis); however, for translation by $L$ along the $x$ direction, a phase shift is created by $\exp{(-ikL)}$, where $k$ is the wave number and $i=\sqrt{-1}$.

The training set for developing ANNs should adequately represent the space of WEC dimensions, including WEC radius and draft.
Since the BEM solution is created in the frequency domain, radial frequency is also an input. 
These inputs are normalized according to the following relationships:
\begin{align}
\tRwec & = \Rwec/\Rwecmax \\
\tDwec & = \Dwec/\Dwecmax\\
\tilde{\omega} & = \omega/\bar{\omega}
\end{align}
\noindent
where $\Rwecmax$, $\Dwecmax$, and $\bar{\omega}$ are maximum WEC radius, draft, and radial frequencies, respectively. 
The input vector for the single-body cluster is then defined as $\tilde{\bm{v}}_{1} = [\tRwec, \tDwec, \tilde{\omega}]^{T}$. 
The QoI for the $1$-body cluster constitutes the normalized added mass $\tilde{\addedmassv}$, damping coefficient $\tilde{\dampingcoeffv}$, and the real and imaginary parts of the excitation force $\tilde{f}_{\text{e}}$.
These quantities are represented as $\tilde{\bm{y}}_{1} = [\tilde{\addedmassv}, \tilde{\dampingcoeffv}, \text{Re}\{\tilde{f}_\text{e}\}, \text{Im}\{\tilde{f}_\text{e}\}]^{T}$.
The resulting surrogate models for $1$-body clusters are then defined as:
\begin{align}
    \tilde{\bm{y}}_{1} = \bm{f}_{1}(\tilde{\bm{v}}_{1})
\end{align}
\noindent
where $\bm{f}_{1}$ is the vector of resulting ANN functions for the single-body cluster, composed of $\bm{f}_{1} = [f^{a}_{1}, f^{b}_{1}, f^{f_{r}}_{1}, f^{f_{im}}_{1} ]^{T}$.
 
The interaction effect among $2$-body clusters is characterized by an input vector that, in addition to WEC radius, draft, and radial frequency, includes the normalized relative distance $\Dispq$ and relative angle $\Angpq$ between the two bodies:
\begin{align}
\tDispq & = \Dispq/\Dispqmax \\
\tAngpq & = \Angpq/\Angpqmax
\end{align}
\noindent
\noindent
where $\Dispqmax$, and $\Angpqmax$ are the maximum distance and angle considered in producing \texttt{Nemoh} results.  
The input vector is then described as $\bm{v}_{2} = [\tRwec, \tDwec, \tDispq, \tAngpq, \tilde{\omega}]^{T}$.   
The radiation output entails elements corresponding to the additive effect described in Eqs.~(\ref{eqn:MBEadditive})--(\ref{eqn:MBEadditive2}).
However, due to numerical reasons, it is more straightforward to structure the two-body surrogate models with direct outputs from \texttt{Nemoh}.
Any required transformation, such as the ones described in Eqs.~(\ref{eqn:MBEadditive})--(\ref{eqn:MBEadditive2}), can be performed once the models are developed. 
The $2$-body system radiation effect depends strictly on the relative distance of the bodies, while the excitation force needs only be investigated in the range of $[0,~\pi]$ because any other angle can be mapped to that solution \cite{zhang2020surrogate}.
These outputs are defined as 
$\tilde{\bm{y}}_{2} = [\tilde{\addedmassv}_{11}, \tilde{\addedmassv}_{12}, \tilde{\dampingcoeffv}_{11},  \tilde{\dampingcoeffv}_{12}, \text{Re}\{\tilde{f}_{\text{e}_{11}}\}, \text{Im}\{\tilde{f}_{\text{e}_{11}}\}]^{T}$, where the indices $\parm_{11}$ and $\parm_{12}$ correspond to the interaction effect between elements of the mass matrix in Eq.~(\ref{eqn:AandBMatrices}) with an array size of $\Nwec = 2$. 
The 2-body cluster surrogate models can be defined as:
\begin{align}
    \tilde{\bm{y}}_{2} = \bm{f}_{2}(\tilde{\bm{v}}_{2})
\end{align}
\noindent
where $\bm{f}_{2}$ is the vector of resulting ANN functions for the two-body cluster, composed of $\bm{f}_{2} = [f^{a_{11}}_{2}, f^{a_{12}}_{2}, f^{b_{11}}_{2},f^{b_{12}}_{2}, f^{f_{r}}_{2}, f^{f_{im}}_{2} ]^{T}$.
From here, the additive effect from the $2$-body cluster associated with the radiation interaction can be defined as:
\begin{align}
    \Delta \tilde{\addedmassv}_{11} & = \tilde{\addedmassv}_{11} - \tilde{\addedmassv} =  f^{a_{11}}_{2}(\tilde{\bm{v}}_{2}) - f^{a}_{1}(\tilde{\bm{v}}_{1}) \label{eq:delta_a_11}\\
    \Delta \tilde{\addedmassv}_{12} & = f^{a_{12}}_{2}(\tilde{\bm{v}}_{2})\label{eq:delta_a_12}\\
    \Delta \tilde{\dampingcoeffv}_{11} &= \tilde{\dampingcoeffv}_{11} - \tilde{\dampingcoeffv} = f^{b_{11}}_{2}(\tilde{\bm{v}}_{2}) - f^{b}_{1}(\tilde{\bm{v}}_{1}) \label{eq:delta_b_11}\\
    \Delta \tilde{\dampingcoeffv}_{12} & = f^{b_{12}}_{2}(\tilde{\bm{v}}_{2}) \label{eq:delta_b_12}
    \end{align}
\noindent
For excitation force, the additive effect is captured as:
\begin{align}
    {\Delta \tilde{f}_{\text{e}11}} &= { (\tilde{f}_{\text{e}} - \tilde{f}_{\text{e}_{11}})\exp{(ikL)}} \label{eq:delta_f_11} \\
    &{= \left ([f^{1}_{fr}(\tilde{\bm{v}}_{1}) + if^{1}_{fim}(\tilde{\bm{v}}_{1})] - [f^{2}_{fr}(\tilde{\bm{v}}_{2}) + if^{2}_{fim}(\tilde{\bm{v}}_{2})]\right)\exp{(ikL)} }\notag
\end{align}
\noindent
Other quantities of interest, such as those corresponding to the additive effect of the first body on the second one, may be simply calculated by swapping the order of the bodies.

\subsection{Data Processing}
\label{subsec:Data_Processing}

In addition to the normalization scheme discussed in Sec.~\ref{subsec:Developing_Surrogate_Models}, further processing of data is necessary in order to improve the performance of ANNs.

Design-informed considerations are utilized in order to generate appropriate inputs for developing ANNs.
While preventing us from producing impractical solutions, such considerations improve the training performance by limiting the range of outputs in the training set.
Here, two practical design decisions are considered when generating the training data. 
First, extreme and unreasonable design combinations are avoided by only considering cases where the radius and draft ratio are within an acceptable range.
This limitation prevents us from generating solutions that have little physical viability. 
Second, as will be described in Sec.~\ref{subsec:Problem_Formulation} Eq.~(\ref{Eqn:distanceconst}), a safety distance, proportional to the radius of the WEC, is necessary for the reliable maintenance of WEC devices.
This safety distance is also considered when generating data for the training of ANNs.

Despite the provisions described above, it was occasionally observed that the data set entailed few extremely high and/or low data points.
This issue might be associated with irregular frequencies in \texttt{Nemoh} solution, which arise due to a fundamental error in BEM formulation \cite{penalba2017using}.    
These data points were identified and replaced by the mean of their neighboring data points.
More advanced methods for removal of irregular frequencies from \texttt{Nemoh} solutions must be investigated in future work \cite{kelly2022post}.

Since the BEM was implemented for a wide range of WEC radii, drafts, distances, and angles, even after normalization, the range of the QoI was extremely high among some of the solution sets.
After some trial and error with various normalization schemes, it was decided that transforming each solution set to the range of $[-1,~1]$ through a linear transformation results in the most satisfactory performance.
However, to do this, additional ANNs need to be developed in order to estimate the range and offset of these linear transformations. 
Through this approach, we ensure that ANNs preserve the shape or profile of the QoI.
This approach resulted in a total of $30$ surrogate models. 

\begin{figure}[t]
    \captionsetup[subfigure]{justification=centering}
    \centering
    \begin{subfigure}{0.5\columnwidth}
    \centering
    \includegraphics[scale=0.5]{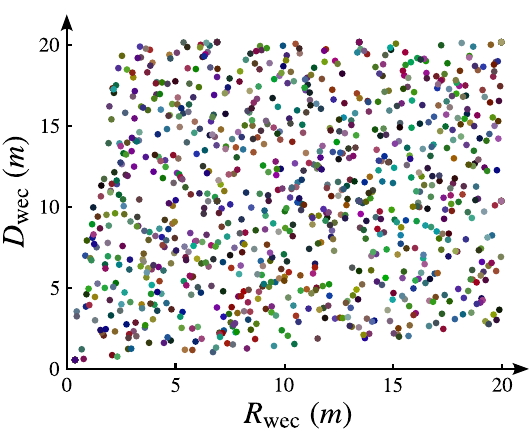}
    \caption{WEC radius and draft.}
    \label{subfig:LHS_distance}
    \end{subfigure}%
    \begin{subfigure}{0.5\columnwidth}
    \centering
    \includegraphics[scale=0.5]{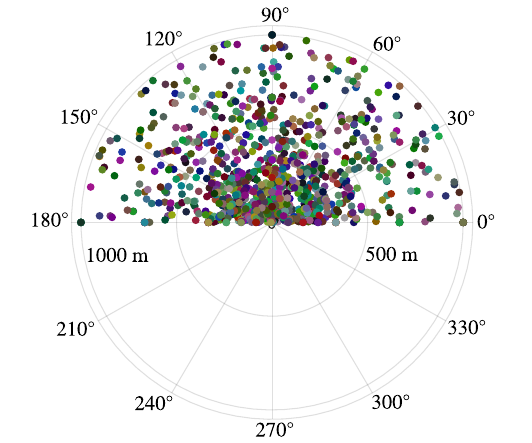}
    \caption{Distance and angle.}
    \label{subfig:LHS_radius}
    \end{subfigure}%
    \captionsetup[figure]{justification=centering}
    \caption{Sampling the design space using LHS for two-body interactions.}
    \label{fig:LHS_sampling}
\end{figure}

\subsection{Validation of Surrogate Models}
\label{subsec:Validation_of_Surrogate_Models}

The BEM solver \texttt{Nemoh} was utilized to create training samples for the development of surrogate models.
A total of $225$ and $1000$ \texttt{Nemoh} calls were made to generate data for the $1$-body- and $2$-body-related surrogate models, respectively.
To improve computational efficiency, customized code was developed to implement \texttt{Nemoh} in parallel, without manipulating \texttt{Nemoh's} source code and executable files.
The resulting package is openly accessible on GitHub upon \cite{codeParNemoh}. 

Uniform spacing was used in creating training data for the $1$-body ANNs. 
However, for the $2$-body ANNs, a Latin-hypercube sampling (LHS) approach was implemented. 
For the distance dimension, this LHS approach was performed using logarithmically-spaced samples. 
The LHS for the two-body ANNs is shown in Fig.~\ref{fig:LHS_sampling}, where each unique color corresponds to an individual data point in the four-dimensional space.

The hydrodynamic coefficients were obtained at the constant water depth of $h=50~\text{m}$ for the range of radial frequencies from $\omega_{0} = 0.1~\text{rad/s}$ to $\omega_{f} = 7~\text{rad/s}$, with a total of $50$ evenly-spaced values.
The mean squared error was consistently used for performance assessment of the ANNs, and the training data was divided such that $70\%$ was used for training, $15\%$ for validation, and $15\%$ for tests.
All the ANNs were trained for up to $30000$ epochs, with early stopping of $10000$ maximum failures in the validation set using \texttt{Matlab}.
General information about generating the training set is presented in Tab.~\ref{Tab:Training_Settings}.

\begin{table}[t]
    \caption{Settings for creating training data using \texttt{Nemoh}.}
    \label{Tab:Training_Settings}
    \renewcommand{\arraystretch}{1.1}
    \centering
    \begin{tabular}{c c c c c}
    \hline  \hline
    \textrm{\textbf{Symbol}} & \textrm{\textbf{Definition}} & \textrm{\textbf{Value}} & \textrm{\textbf{Unit}} \\
    \hline
    $\omega$ & \textrm{frequency} & $[0.1~~ 7]$ & \textrm{rad/s} \\
    $h$ & \textrm{water depth}&$50$ & \textrm{m} \\
    $\beta_{w}$ & \textrm{wave angle}& $0$ & \textrm{rad} \\
    $\Rwec$ & WEC radius & $[0.5,~20]$ & \textrm{m} \\
    $\Dwec$ & WEC draft & $[0.5,~20]$ & \textrm{m} \\
    $\Dispq$ &  Relative distance & $[2\Rwec + s_d,~1000]$ & \textrm{m}\\
    $\Angpq$ &  Relative angle & $[0,~\pi]$ & \textrm{rad} \\
    $n_{s_{1}}$ & \textrm{\# $1$-body samples} & $225$ & -\\
    $n_{s_{2}}$ & \textrm{\# $2$-body samples} & $1000$ & - \\
    \hline \hline
    \end{tabular}
\end{table}

\begin{figure*}[t]
\centering
\includegraphics[scale = 0.6]{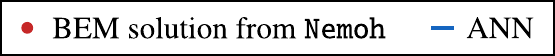}\\
\captionsetup[subfigure]{justification=centering}
\centering
\begin{subfigure}{0.25\textwidth}
\centering
\includegraphics[scale=0.595]{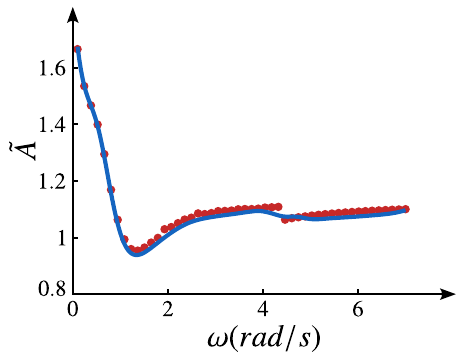}
\caption{Added mass.}
\label{fig:A_1wec_validation}
\end{subfigure}%
\begin{subfigure}{0.25\textwidth}
\centering
\includegraphics[scale=0.595]{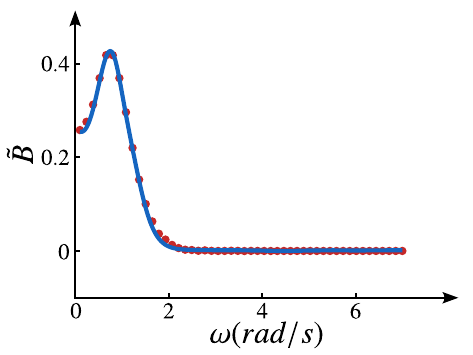}
\caption{Damping coefficient.}
\label{fig:B_1wec_validation}
\end{subfigure}%
\begin{subfigure}{0.25\textwidth}
\centering
\includegraphics[scale=0.595]{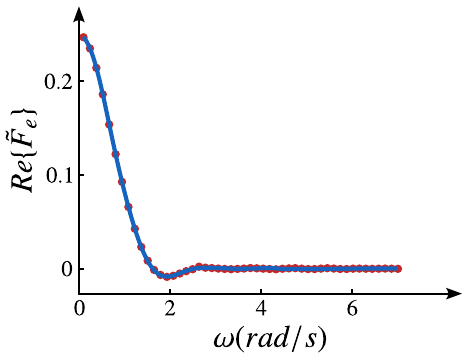}
\caption{Excitation force (real).}
\label{fig:Fer_1wec_validation}
\end{subfigure}%
\begin{subfigure}{0.25\textwidth}
\centering
\includegraphics[scale=0.595]{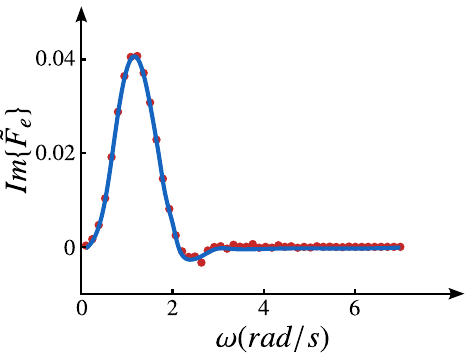}
\caption{Excitation force (imaginary).}
\label{fig:Feim_1wec_validation}
\end{subfigure}%
\captionsetup[figure]{justification=centering}
\caption{Validation of ANNs with BEM solution from \texttt{Nemoh} using $\Rwec = 8~\textrm{m}$ and $\Dwec = 4~\textrm{m}$.} 
\label{fig:1wec_validation}
\end{figure*}

\begin{figure*}[t]
\centering
\includegraphics[scale = 0.6]{CommonLegend.pdf}\\
\captionsetup[subfigure]{justification=centering}
\centering
\begin{subfigure}{0.25\textwidth}
\centering
\includegraphics[scale=0.595]{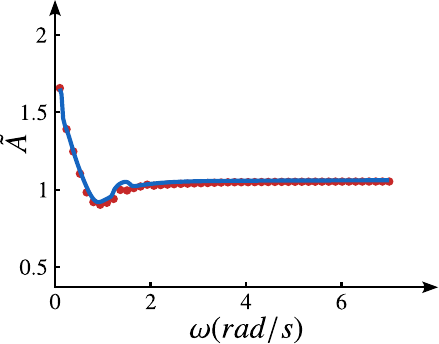}
\caption{Added mass.}
\label{fig:A11_2wec_validation}
\end{subfigure}%
\begin{subfigure}{0.25\textwidth}
\centering
\includegraphics[scale=0.595]{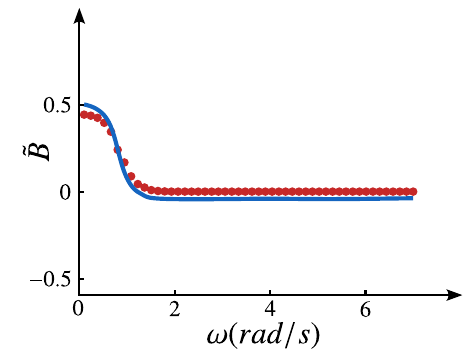}
\caption{Damping coefficient.}
\label{fig:A12_2wec_validation}
\end{subfigure}%
\begin{subfigure}{0.25\textwidth}
\centering
\includegraphics[scale=0.595]{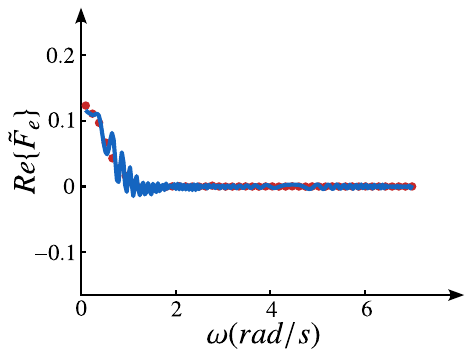}
\caption{Excitation force (real).}
\label{fig:B11_2wec_validation}
\end{subfigure}%
\begin{subfigure}{0.25\textwidth}
\centering
\includegraphics[scale=0.595]{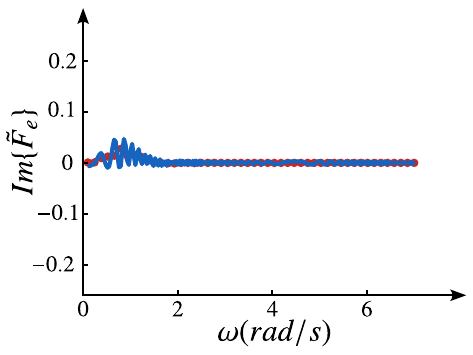}
\caption{Excitation force (imaginary).}
\label{fig:B12_2wec_validation}
\end{subfigure}%
\captionsetup[figure]{justification=centering}
\caption{Validation of select ANNs with BEM solution from Nemoh using $\Rwec = 8~\textrm{m}$, $\Dwec = 15~\textrm{m}$, $\Dispq = 200~\textrm{m}$, and $\Angpq =0.078~\textrm{radians}$.} 
\label{fig:SM_2WEC_Validate}
\end{figure*}

The resulting ANNs are first validated against arbitrary design specifications using \texttt{Nemoh} to ensure reasonable performance.
Using $\Rwec = 8~\textrm{m}$ and $\Dwec = 4~\textrm{m}$, the validation plots for added mass, damping coefficient, and real and imaginary parts of the excitation force for the $1$-body cluster are presented in Fig.~\ref{fig:1wec_validation}, where it is clear that the ANNs are in reasonable agreement with \texttt{Nemoh}.
Similarly, for the two-body interaction effect, the developed ANNs were compared with \texttt{Nemoh} results for an arbitrary case of $\Rwec = 15~\textrm{m}$, $\Dwec = 8~\textrm{m}$, $\Dispq = 200~\textrm{m}$, and $\Angpq = 0.078~\textrm{radians}$.
The comparison plots for the added mass and damping coefficients and the real and imaginary parts of the excitation forces are shown in Fig.~\ref{fig:SM_2WEC_Validate}.

\xsection{Concurrent UCCD and Layout Optimization}
\label{sec:UCCDandLayoutOpt}

Optimal spacing of various WEC technologies within an array and optimal sizing and control of WEC devices have been popular topics in the literature, often investigated individually \cite{sergiienko2021effect, esmaeilzadeh2019shape, mcguinness2017constrained}.
The significant impact of these elements on WEC performance has convinced more researchers to concurrently investigate such attributes of these complex systems.
For example, Ref.~\cite{lyu2019optimization} carries out a concurrent sizing and layout optimization of WECs by taking advantage of the simplified $q$-factor equation with the point-absorber approximation.
Various optimization methods have been used for the layout optimization of WECs, including sequential quadratic programming \cite{fitzgerald2007preliminary}, genetic algorithm (GA) \cite{tay2017optimization}, and heuristics \cite{moarefdoost2017layouts}. 
This section presents the simultaneous formulation and some preliminary results for the concurrent plant, control, and layout optimization of WEC farms.

\subsection{Problem Formulation}
\label{subsec:Problem_Formulation}
Since manufacturing costs associated with WECs of different dimensions become prohibitive, it is reasonable to assume that WEC dimensions are uniform across the farm.
However, to maximize energy generation, control parameters associated with the PTO system of each WEC, along with their location (thus, the configuration of the farm), may be varied.
Expressing the objective function as the average power per unit volume of the device, the concurrent UCCD and layout optimization problem can be formulated as:
\begin{subequations}
 \label{Eqn:OPtimization}
 \begin{align}
 \underset{\bm{p},\bm{u}, \AL}{\textrm{minimize:}}
 \quad & - p_{v}(\bm{p}, \bm{u}, \AL)   \label{Eqn:Obj} \\
 \textrm{subject to:} \quad
    \begin{split}
    &  2\Rwec + s_{d} - \bm{L}_{pq} \leq 0 \quad\\
         & \qquad \forall ~~  p,q = 1, 2, \dots, \Nwec \quad p \neq q \label{Eqn:distanceconst}
    \end{split} \\
    &  \underaccent{\bar}{\bm{p}} \leq \bm{p} \leq \bar{\bm{p}}\label{Eqn:plantconst}\\
    &  \underaccent{\bar}{\bm{u}} \leq \bm{u} \leq \bar{\bm{u}} \label{Eqn:controlconst}\\
    &  \underaccent{\bar}{\AL} \leq \AL \leq \bar{\AL}\label{Eqn:layoutconst}\\
 \textrm{where:} \quad & \bm{p} = [\Rwec, \Dwec]^{T} \in \mathbb{R}^{2} \notag\\
                & \bm{u} = [\KPTO, \BPTO]^{T} \in \mathbb{R}^{{2}\Nwec}  \notag \\
                & \AL = [\bm{x},\bm{y}] \in \mathbb{R}^{{2}(\Nwec-1)} \notag 
 \end{align} 
\end{subequations}
\noindent
where $\bm{p}$ is the vector of plant optimization variables composed of WEC radius $\Rwec$ and draft $\Dwec$ (uniform across the farm). 
$\bm{u}$ is the vector of time-independent control parameters for each of the WEC devices, composed of PTO spring stiffness $\KPTO$ and damping $\BPTO$.
The layout design vector $\AL$ is composed of $(\Nwec - 1)$ coordinates, corresponding to the $x$- and $y$-axis locations of WEC devices.
Note that here, the first WEC is always assumed to be positioned at the center of the coordinate system and, thus, is not included in $\AL$. 
Equation (\ref{Eqn:distanceconst}) ensures that the minimum distance between each pair of WECs is larger than the WEC diameter plus an additional safe distance $s_{d}$, which is required to allow maintenance ships to pass safely \cite{neshat2022layout}.
Here, we calculate the safe distance as a function of $\Rwec$, such that $s_{d} = (\Rwec/5)\times50~m$.
All of the optimization variables are confined within their associated bounds, as described by Eqs.~(\ref{Eqn:plantconst})--(\ref{Eqn:layoutconst}).
The farm area is restricted to a box with dimensions of $\pm 0.5 \times\sqrt{20000\Nwec}~\textrm{m}$ in both $x$ and $y$ axes \cite{neshat2022layout}.
These details, along with some additional problem data, are presented in Table~\ref{Tab:Parameter_Opt}.

 \begin{table}[t]
    \caption{Problem parameters.}
    \label{Tab:Parameter_Opt}
    \renewcommand{\arraystretch}{1.1}
    \setlength{\tabcolsep}{4pt}
    \centering
    \begin{tabular}{s c s c}
    \hline  \hline
    \textrm{\textbf{Option}} & \textrm{\textbf{Value}} &  \textrm{\textbf{Option}} & \textrm{\textbf{Value}} \\
    \hline
    $\Rwecmin$ & $0.5~\textrm{m}$ & $\Rwecmax$ & $10~\textrm{m}$ \\
    $\Dwecmin$ & $0.5~\textrm{m}$ & $\Dwecmax$ & $10~\textrm{m}$ \\
    $\KPTOmin$ & $-3\times 10^{8}~\textrm{N/m}$ & $\KPTOmax$ & $3\times 10^{8}~\textrm{N/m}$ \\
    $\BPTOmin$ & $0~ \textrm{Ns/m}$ & $\BPTOmax$ & $3\times 10^{8}~ \textrm{Ns/m}$\\
    $\underaccent{\bar}{\bm{x}}$ & $-0.5\sqrt{2\Nwec \times 10^4}~\textrm{m}$ &  $\bar{\bm{x}}$ & $0.5\sqrt{2\Nwec \times 10^4}~\textrm{m}$ \\
    $\underaccent{\bar}{\bm{y}}$ & $-0.5\sqrt{2\Nwec \times 10^4}~\textrm{m}$ &  $\bar{\bm{y}}$ & $0.5\sqrt{2\Nwec \times 10^4}~\textrm{m}$ \\
    $\rho$ & $1025~ \textrm{kg/m$^3$}$ & $g$ & $9.81~\textrm{m/s$^{2}$}$ \\
    $s_{d}$ & $50 \times {\Rwec/5}~\textrm{m}$ & $\Nwec$ & $[3, 5, 7, 10]$ \\
    $n_{yr}$   & $30~\textrm{years}$ & $n_{r}$  & $200$\\
    $n_{gq}$   & $20$  & $\eta_{\text{pcc}}$ & $0.8$ \\
    $\eta_{\text{oa}}$ & $0.95$ & $\eta_{\text{t}}$ & $0.98$\\
    \hline \hline
    \end{tabular}
\end{table}

\subsection{Results and Discussion}
\label{subsec:Results_and_Discussion}

\begin{figure*}[t]
\centering
\includegraphics[scale = 0.8]{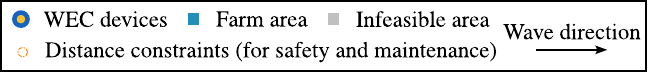}\\
\captionsetup[subfigure]{justification=centering}
\centering
\begin{subfigure}{0.25\textwidth}
\centering
\includegraphics[scale=0.595]{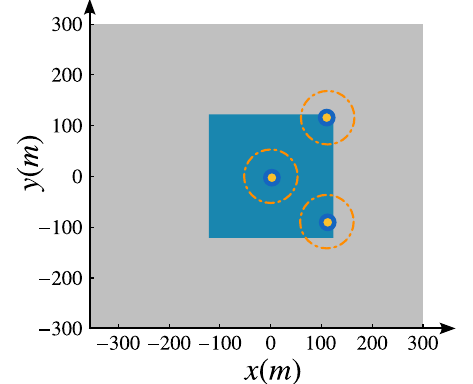}
\caption{3-WEC optimized array.}
\label{fig:3-WEC}
\end{subfigure}%
\begin{subfigure}{0.25\textwidth}
\centering
\includegraphics[scale=0.595]{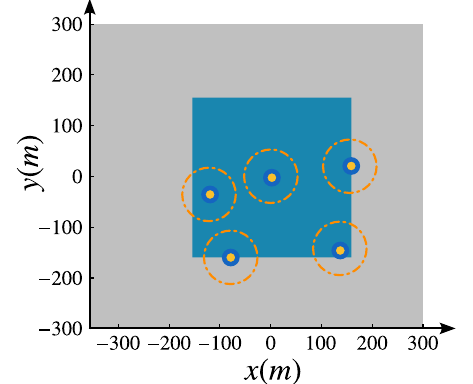}
\caption{5-WEC optimized array.}
\label{fig:5-WEC}
\end{subfigure}%
\begin{subfigure}{0.25\textwidth}
\centering
\includegraphics[scale=0.595]{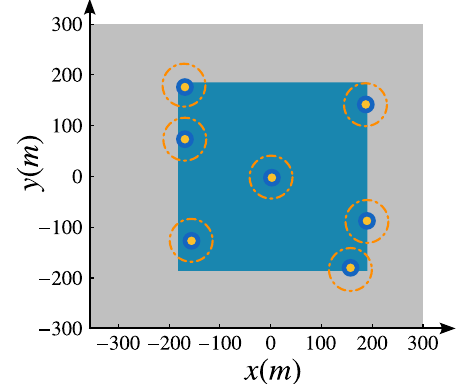}
\caption{7-WEC optimized array.}
\label{fig:7-WEC}
\end{subfigure}%
\begin{subfigure}{0.25\textwidth}
\centering
\includegraphics[scale=0.595]{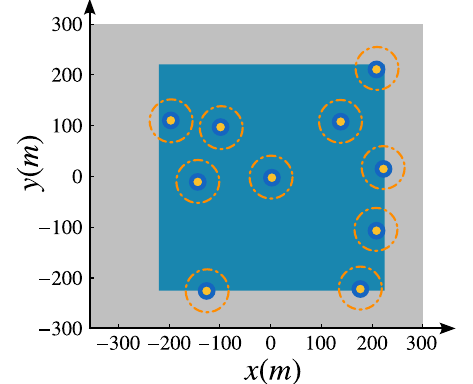}
\caption{10-WEC optimized array.}
\label{fig:10-WEC}
\end{subfigure}%
\captionsetup[figure]{justification=centering}
\caption{Optimized array layout with farm area and distant constraints.} 
\label{fig:OptimalArray}
\end{figure*}

The concurrent plant, control, and layout optimization problem was implemented for four case studies with $\Nwec = 3, 5, 7$, and $10$.
Due to the complex nature of the problem, global optimization is required.
Therefore, \texttt{Matlab}'s \textit{surrogateopt} solver, specifically developed for the global optimization of time-consuming objective functions, was used for these case studies and ran for a total of $300$ function evaluations using parallel computation.
The computer architecture used for these case studies is a desktop workstation with an AMD Ryzen Threadripper 3970X 32-core processor at 3.69 \unit{GHz}, 128 \unit{GB} of RAM, 64-bit windows 10 Enterprise LTSC version 1809, and \texttt{Matlab} R2022a. 
The results are presented in Table~\ref{Tab:results}.

\begin{table*}[t]
    \caption{Concurrent UCCD and layout optimization solutions.}
    \label{Tab:results}
    \renewcommand{\arraystretch}{1.1}
    \centering
    \begin{tabular}{c c c c c c c c}
    \hline  \hline
    \textrm{\textbf{Case Study}} & $\Rwec~\textrm{[m]}$ &  $\Dwec~\textrm{[m]}$ &  $\BPTO~\textrm{[Ns/m]}$ & $\KPTO~\textrm{[N/m]}$ & $\AL~\textrm{[m]}$ & $p_{v}~\textrm{[MW/m$^3$]}$ & \textrm{ Time~\textrm{[s]}} \\
    \hline
    \multirow{3}{*}{3-WEC} & \multirow{3}{*}{$8.83$} & \multirow{3}{*}{$0.54$} & $1.7 \times 10^8$ & $-2.97 \times 10^8$ & $[0,0]$ & \multirow{3}{*}{$175.58$} & \multirow{3}{*}{$97$} \\
    &  &  & $1.65 \times 10^8$ & $2.53 \times 10^8$ & $[109, -89.52]$ &    &  \\
    &  &  & $1.1 \times 10^8$ & $-4.21 \times 10^6$ & $[114.07,115.89]$ &   & \\
    \hline 
    \multirow{5}{*}{5-WEC} & \multirow{5}{*}{$7.12$} &  \multirow{5}{*}{$0.5$} & $2.64 \times 10^8$  & $1.44 \times 10^8$  & $[0,0]$ & \multirow{5}{*}{$566.02$}  & \multirow{5}{*}{$143$}\\
    &  &  & $1.67 \times 10^8$  & $5.26 \times 10^6$  & $[-138.76, -119.09]$ &    & \\
    &  &  & $1.67 \times 10^8$  & $2.56 \times 10^8$  & $[-24.07, -154.43]$ &    & \\
    &  &  & $2.94 \times 10^8$  & $-2.95 \times 10^8$  & $[-41.54, 129.69]$ &   & \\
    &  &  & $2.72 \times 10^8$  & $2.73 \times 10^8$  & $[-49.97, -70.61]$ &   & \\
    \hline
    \multirow{7}{*}{7-WEC} & \multirow{7}{*}{$6.71$} & \multirow{7}{*}{$0.62$} & $8.86 \times 10^7$ & $-2.95 \times 10^8$ & $[0,0]$ & \multirow{7}{*}{$2.21 \times 10^3$} & \multirow{7}{*}{$219$}\\
    &  &  & $1.94 \times 10^8$ & $-3.24 \times 10^7$ & $[154,-181.38]$ &    & \\
    &  &  & $1.16 \times 10^8$ & $1.32 \times 10^8$ & $[-169.51,73.9]$ &    & \\
    &  &  & $6.66 \times 10^7$ & $-8.66 \times 10^6$ & $[-171.96,178.79]$ &  & \\
    &  &  & $2.84 \times 10^8$ & $2.32 \times 10^8$ & $[185.47,142.23]$ &   & \\
    &  &  & $2.67 \times 10^8$ & $1.69 \times 10^8$ & $[187.08,-86.12]$ &  & \\
    &  &  & $2.67 \times 10^8$ & $4.4 \times 10^7$ & $[-156.86,-124.98]$ &  & \\
    \hline
    \multirow{10}{*}{10-WEC} &  \multirow{10}{*}{$7.32$} &  \multirow{10}{*}{$0.5$} & $2.24 \times 10^8$ & $-1.73\times 10^8$ & $[0,0]$ & \multirow{10}{*}{$4.12 \times 10^3$} & \multirow{10}{*}{$379$} \\
    &  &  & $2.99 \times 10^8$ & $-2.56 \times 10^7$ & $[-100.66, 98.45]$ &   & \\
    &  &  & $1.29 \times 10^8$ & $-2.83 \times 10^8$ & $[-127.97, -223.61]$ &    & \\
    &  &  & $6.9 \times 10^7$ & $1.82 \times 10^8$ & $[-145.75, 8.36]$ &    & \\
    &  &  & $2.41 \times 10^8$ & $-2.08 \times 10^7$ & $[219.94, 18.84]$ &    &\\
    &  &  & $2.23 \times 10^8$ & $-6.74 \times 10^7$ & $[-198.02, 112.55]$ &    & \\
    &  &  & $1.89 \times 10^8$ & $2.14 \times 10^7$ & $[206.39,214.89 ]$ &  &   \\
    &  &  & $2.68 \times 10^7$ & $1.45 \times 10^8$ & $[204.82, -102.91]$ &    & \\
    &  &  & $1.81 \times 10^8$ & $-2.27 \times 10^8$ & $[174.45, -218.4]$ &  &   \\
    &  &  & $1.66 \times 10^8$ & $1.49 \times 10^7$ & $[135.33, 109.09]$ &    & \\
    \hline \hline
    \end{tabular}
\end{table*}

According to this table, the proposed framework enables the concurrent control co-design and layout optimization of WEC farms with significant improvements in computational efficiency, making it possible to investigate large arrays within a reasonable amount of time.
This improvement in efficiency is signified by the fact that with the proposed approach, a $10$-WEC array optimization study can be performed much faster than a single \texttt{Nemoh} call to solve a $3$-WEC hydrodynamic problem.  
An expected tradeoff is the accuracy of the estimations, which is mainly affected by two contributions, one from overlooking $3$ and higher-body interactions in developing MBE models and the other from inherent errors associated with the development of surrogate models.   
Upon further improvement of the accuracy of these estimations, the proposed method may offer more practical, early-stage insights into the interactions between the design, control, and layout of these complex devices.

The optimized layout configurations associated with all of the studies are described in Table~\ref{Tab:results} and visualized along with farm area and distance constraints in Fig.~\ref{fig:OptimalArray}.
The table shows that the radius of WEC devices is relatively high (about $8.83~\text{m}$) for $3$-WEC farm, while it decreases in size for the other studies.  
This result seems to be consistent with some of the results from the literature, where it is shown that multiple smaller devices result in a higher power \cite{o2013techno, de2015adaptability, de2016techno}.
However, complete validation of such literature results requires the optimizer to determine the number of WEC devices, which can be a future direction of this research.  
The draft dimensions in all of these studies remain very close to the lower bound of $0.5~\text{m}$. 
This result is also consistent with the results from the literature, where it is shown that the draft size does not have a significant impact on power generation \cite{khojasteh2016evaluation}.

The variations in $\BPTO$ and $\KPTO$ are much more diverse, indicating that individual control of WECs is necessary for improved device performance.
Here, we have allowed the spring stiffness to take negative values.
These stiffness results enable a reactive phase control strategy \cite{antonio2010wave} that is often used to extend the range of resonance conditions in WEC design \cite{todalshaug2016tank, peretta2015effect}.
Despite the limitations of this approach (as stated in  Ref.~\cite{antonio2010wave}), it is shown that phase control with a negative spring stiffness increases the absorbed power \cite{todalshaug2016tank}. 
In each case study, the PTO damping is selected by the optimizer in order to maximize the energy production per volume of the WEC device over the lifetime of the farm. 
This approach considers the availability of wave resources when designing the PTO damping coefficient. 

The optimized layout configurations and their associated distance constraints are presented in Fig.~\ref{fig:OptimalArray}. 
While Ref.~\cite{moarefdoost2017layouts} has shown that, for the most part, the best layout arrays are symmetric with respect to the wave direction, this symmetry only occurs for the $3$-WEC study in Fig.~\ref{fig:3-WEC}.
This outcome might be mainly associated with \textit{(i)} the need to further improve the estimation of the hydrodynamic interaction effect by accounting for higher-order terms in MBE, \textit{(ii)} improving upon the accuracy of the lower-order surrogate models, and \textit{(iii)} allowing the optimizer to run for a higher number of function evaluations.
Additionally, it is also possible that the influence of the separating distance between WECs is limited due to proper tuning of the PTO damping coefficient \cite{borgarino2012impact}.

The generated power per unit volume of the device is also shown in the table.
It is evident that the power generation per unit volume of the device, as well as the overall power generation (not shown here), increases as we increase the number of devices in the farm.  
The average amount of power generated by an individual device per year in each layout also increases as we increase the number of devices in the farm. 

While the results presented here offer some insights into the effectiveness of the proposed approach in terms of computational expense by utilizing MBE and surrogate modeling, conclusions regarding trends in array configurations and the amount of generated power are avoided at this stage.
This decision is because the accuracy tradeoff in developing surrogate models must be further investigated, and higher-order terms must be included in the MBE approach in order to capture the complex interactions in the farm better. 
In addition, this study considers no limits on PTO force and buoy displacement. 
This assumption implies that the device can generate power from waves with very high amplitudes without considering any of the physical, structural, and operational constraints.
For these reasons, we avoid making any recommendations regarding optimal layout configurations but emphasize that the proposed approach and methodology have the potential to address the computational bottleneck in the concurrent plant, control, and layout optimization of large arrays.

\xsection{Conclusion}\label{sec:conclusion}

Due to nonlinear and complex dynamics, the sizing, control, and array layout optimization of wave energy converters (WECs) are coupled disciplines and must be approached concurrently from the early stages of the design process.
However, estimating the hydrodynamic interaction effect for this integrated design study is computationally complex.
Therefore, in this article, we developed data-driven artificial neural networks (ANNs) and utilized the principles of many-body expansion (MBE) to efficiently calculate the hydrodynamic interaction effect up to the second-order terms.
The heaving WEC model was developed in the frequency domain, with WEC radius and draft as plant optimization variables that were assumed to be uniform across the farm. 
The power take-off control damping, stiffness parameters, and array layout were optimized for each single WEC device using a global optimization algorithm. 
The results indicated significant computational efficiency compared to the direct usage of boundary element method solvers in the optimizer loop.

While the current study shows promising directions for efficiently estimating the hydrodynamic interaction effect, larger array studies necessitate using higher-order terms in the MBE equation.  
In addition, the variations in water depth and geographical location of the farm must be considered to make general recommendations. 
More complex objective functions that consider techno-economic-related metrics involving costs for manufacturing, operation, PTO, maintenance, etc., must be investigated. 
Finally, force saturation limits are necessary to ensure practical infrastructure requirements and must be considered as additional constraints in the optimization problem for future work.

\begin{acknowledgment}
The authors gratefully acknowledge the financial support from National Science Foundation Engineering Design and Systems Engineering Program, USA under grant number CMMI-2034040.
We would also like to thank Gaofeng Jia and Akshat Chulahwat for their contributions.
\end{acknowledgment}

\renewcommand{\refname}{REFERENCES}
\bibliographystyle{asmems4}
{\small
\bibliography{References}
}

\nomenclature[A, 01]{\(\textrm{ANN}\)}{artificial neural network}
\nomenclature[A, 02]{\(\textrm{BEM}\)}{boundary element method}
\nomenclature[A, 03]{\(\textrm{CCD}\)}{control co-design}
\nomenclature[A, 04]{\(\textrm{CFD}\)}{computational fluid dynamics}
\nomenclature[A, 05]{\(\textrm{LHS}\)}{Latin hypercube sampling}
\nomenclature[A, 06]{\(\textrm{MBE}\)}{many body expansion}
\nomenclature[A, 07]{\(\textrm{PTO}\)}{power take-off}
\nomenclature[A, 08]{\(\textrm{QoI}\)}{quantities of interest}
\nomenclature[A, 09]{\(\textrm{SE-UCCD}\)}{stochastic in expectation uncertain control co-design}
\nomenclature[A, 10]{\(\textrm{TRL}\)}{technology readiness level}
\nomenclature[A, 11]{\(\textrm{WEC}\)}{wave energy converter}

\nomenclature[v, 01]{\( \AddedMass \)}{added mass matrix}
\nomenclature[v, 02]{\( \DampingCoeff \)}{damping coefficient matrix}
\nomenclature[v, 03]{\( B_{\text{pto}} \)}{PTO damping}
\nomenclature[v, 04]{\( \Dwec \)}{WEC draft}
\nomenclature[v, 05]{\( \hat{\Force}_{\text{FK}} \)}{Froude-Krylov force}
\nomenclature[v, 06]{\( \hat{\Force}_{\text{s}} \)}{scattering force}
\nomenclature[v, 07]{\( \hat{\Force}_{\text{r}} \)}{radiation force}
\nomenclature[v, 08]{\( \hat{\Force}_{\text{hs}} \)}{hydrostatic force}
\nomenclature[v, 09]{\( \hat{\Force}_{\text{pto}} \)}{PTO force}
\nomenclature[v, 10]{\( \hat{\Force}_{\text{e}} \)}{excitation force}
\nomenclature[v, 09]{\( G \)}{hydrostatic coefficient}
\nomenclature[v, 11]{\( H_{s}\)}{significant wave height}
\nomenclature[v, 12]{\( k_{\text{pto}} \)}{PTO stiffness}
\nomenclature[v, 13]{\( \Mass \)}{mass matrix}
\nomenclature[v, 14]{\( \Rwec \)}{WEC radius}
\nomenclature[v, 15]{\( S \)}{cross-sectional area at undisturbed sea level}
\nomenclature[v, 16]{\( S_{JS}(\cdot)\)}{JONSWAP spectrum}
\nomenclature[v, 17]{\( T_{p} \)}{wave period}
\nomenclature[v, 18]{\( \mathbf{W} \)}{layout matrix}

\nomenclature[v, 19]{\( f_{1} \)}{first-order ANN function vector}
\nomenclature[v, 20]{\( f_{2} \)}{second-order ANN function vector}
\nomenclature[v, 21]{\( h \)}{water depth}
\nomenclature[v, 22]{\( \Dispq \)}{relative distance between $p$th and $q$th WEC}
\nomenclature[v, 23]{\( \NMBE \)}{order of MBE}
\nomenclature[v, 24]{\( n_{gq} \)}{number of Gauss quadrature points}
\nomenclature[v, 25]{\( n_{yr} \)}{life of the farm}
\nomenclature[v, 26]{\( p_{v} \)}{power per unit volume}
\nomenclature[v, 27]{\( \bm{p}_{m} \)}{time-averaged absorbed mechanical power }
\nomenclature[v, 28]{\( \tilde{v}_{1} \)}{ANN input for 1-body cluster}
\nomenclature[v, 29]{\( \tilde{v}_{2} \)}{ANN input for 2-body cluster}
\nomenclature[v, 30]{\( \bm{w}_{i} \)}{layout vector for $i$th WEC}
\nomenclature[v, 31]{\( \tilde{y}_{1} \)}{ANN output for 1-body cluster}
\nomenclature[v, 32]{\( \tilde{y}_{2} \)}{ANN output for 2-body cluster}

\nomenclature[v, 33]{\( \alpha \)}{spectrum parameter}
\nomenclature[v, 34]{\( \beta \)}{spectrum parameter}
\nomenclature[v, 35]{\( \beta_{w} \)}{wave angle}
\nomenclature[v, 36]{\( \Delta\psi \)}{additive hydrodynamic effect}
\nomenclature[v, 37]{\( \Angpq \)}{relative angle between $p$th and $q$th WEC}
\nomenclature[v, 38]{\( \hat{\bm{\xi}} \)}{displacement matrix}
\nomenclature[v, 39]{\( \rho \)}{water density}
\nomenclature[v, 40]{\( \psi \)}{hydrodynamic effect}
\nomenclature[v, 41]{\( \omega \)}{radial frequency}

\printnomenclature

\end{document}